\documentclass[twocolumn,preprintnumbers,amsmath,amssymb]{revtex4}
\usepackage{epsfig}
\begin{document}

\title{Spin-torque driven ferromagnetic resonance \\ of Co/Ni
synthetic layers in spin valves}

\author{W. Chen, J-M. L. Beaujour, G. de Loubens, A. D. Kent}
\affiliation{Department of Physics, New York University, New York,
NY 10003}
\author{J. Z. Sun}
\affiliation{IBM T. J. Watson Research Center, Yorktown Heights, NY
10598}
\date{November 30th, 2007}

\begin{abstract}
Spin-torque driven ferromagnetic resonance (ST-FMR) is used to study
thin Co/Ni synthetic layers with perpendicular anisotropy confined
in spin-valve based nanojunctions. Field swept ST-FMR measurements
were conducted with a magnetic field applied perpendicular to the
layer surface. The resonance lines were measured under low amplitude
rf excitation, from 1 to 20 GHz. These results are compared with
those obtained using conventional rf field driven FMR on extended
films with the same Co/Ni layer structure. The layers confined in
spin valves have a lower resonance field, a narrower resonance
linewidth and approximately the same linewidth \textit{vs} frequency
slope, implying the same damping parameter. The critical current for
magnetic excitations is determined from measurements of the
resonance linewidth \textit{vs} dc current and is in accord with the
one determined from I-V measurements.
\end{abstract}
\maketitle

Spin-transfer torque has been theoretically predicted and
experimentally demonstrated to drive magnetic excitations in
nanostructured spin valves and magnetic tunnel junctions
\cite{Slonczewski1996, Berger1996, Katine2000, Sun1999, Huai2004}.
With an rf current, spin transfer can be used to study ferromagnetic
resonance \cite{Tulapurkar2005, Sankey2006}. This technique, known
as spin-torque driven ferromagnetic resonance (ST-FMR), enables
quantitative studies of the magnetic properties of thin layers in a
spin-transfer device. Specifically, the layer magnetic anisotropy
and damping can be determined \cite{Fuchs2007}, which are important
parameters that need to be optimized in spin-torque-based memory and
rf oscillator applications.

Spin-transfer memory devices will likely include magnetic layers
with perpendicular magnetic anisotropy that counteracts their
shape-induced easy-plane anisotropy. This will allow efficient use
of spin current for magnetic reversal with a reduced switching
threshold \cite{Sun2000} and a faster switching process
\cite{Kent2004}. Recent work by Mangin \textit{et al.}
\cite{Fullerton2006} has demonstrated improvements of spin-torque
efficiency in a spin valve that has perpendicularly magnetized Co/Ni
synthetic layers. For further optimization of perpendicular
anisotropy materials, it is important to have quantitative
measurements of their anisotropy field and damping in a
nanostructured device, as both of these parameters directly affect
the threshold current for spin-transfer induced switching.

In this Letter, we present ST-FMR studies of bilayer nanopillars,
where the thin (free) layer is composed of a Co/Ni synthetic layer
and the thick (fixed) layer is pure Co. The magnetic anisotropy and
damping of the Co/Ni have been determined by ST-FMR. We compare
these results with those obtained from extended films with the same
Co/Ni layer stack measured using traditional rf field driven FMR.

\begin{figure}[t]
\includegraphics[width=0.48\textwidth]{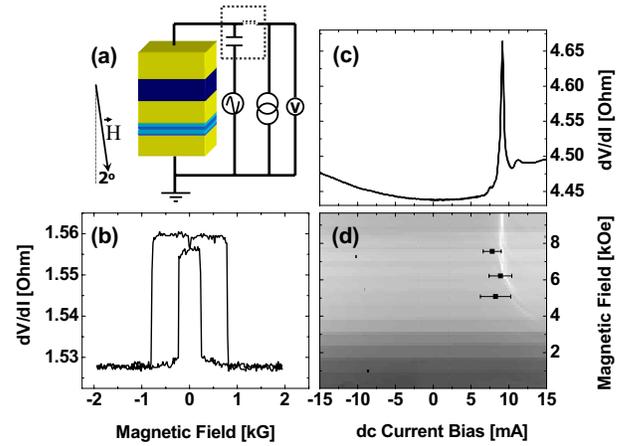}
\caption{(a): Sample layer structure and ST-FMR circuit. (b): Zero
current in-plane MR hysteresis loop for a 50$\times$150 nm$^{2}$
spin valve junction with \textit{t}=0.4. (c): $dV/dI$ \textit{vs} I
of the same junction with a perpendicular magnetic field of 9.5 kOe.
(d): Contour plot of $dV/dI$ as a function of both dc current and
perpendicular magnetic field. Data points: critical currents
determined from ST-FMR at three different fields and frequencies
(see text).} \label{DC}
\end{figure}

Pillar junctions with submicron lateral dimensions (Fig.
\ref{DC}(a)) were patterned on a silicon wafer using a nanostencil
process \cite{Sun2002}. Junctions were deposited using metal
evaporation with the layer structure $\parallel1.5$ nm Cr$\mid 100$
nm Cu$\mid 20$ nm Pt$\mid 10$ nm Cu$\mid$ [\textit{t} nm Co$\mid$
2\textit{t} nm Ni]$\times$ 1.2/\textit{t} $\mid 10$ nm Cu$\mid 12$
nm Co$\mid 200$ nm Cu$\parallel$. We varied the Co thickness
\textit{t} from 0.1 to 0.4, tuning the magnitude of the Co/Ni
composite layer's net anisotropy, while keeping the total magnetic
moment and thickness of the free layer constant. For ST-FMR
measurements, an rf current generated by a high frequency source is
added to a dc current using a bias-T (the dashed-line box in Fig.
\ref{DC}(a)). Positive dc currents are defined such that electrons
flow from the free layer to the fixed layer.

The magnetoresistance (MR) was measured with a magnetic field
applied in the film plane using a 4-point geometry. A typical MR
hysteresis loop of a 50$\times$150 nm$^{2}$ junction with
\textit{t}=0.4 is shown in Fig. \ref{DC}(b). The magnetoresistance
MR$=(R_{AP}-R_{P})/R_{P}$ is $\simeq$2.3$\pm$0.3 $\%$ for all
junctions, independent of \textit{t}, within the range investigated.
Here $R_{AP}$ ($R_{P}$) represents the static junction resistance
when the free layer and fixed layer magnetizations are antiparallel
(parallel). Current-voltage measurements were conducted with a
magnetic field applied \textit{nearly} perpendicular to the sample
surface (The field was applied $2^{\circ}$ from the film normal to
produce a small in-plane field along the easy axis of the junction.
This was done to suppress vortex states in the magnetic layers.)
Measurements were conducted in a 2-point geometry where lead
resistances are included. Fig. \ref{DC}(c) shows $dV/dI$ \textit{vs}
I of the same junction in a 9.5 kOe applied field. A peak without
hysteresis is observed at 9.1 mA, which we interpret as the critical
current $I_{c}$ for excitation of the free layer
\cite{Ozyilmaz2003}. A contour plot of 2-point $dV/dI$ as the
function of both current and perpendicular magnetic field is shown
in Fig. \ref{DC}(d). The peak in $dV/dI$ is seen as the bright color
at high field and current.

At resonance, the rf current and spin valve resistance oscillate at
the same frequency resulting in a dc voltage ($V=<I(t)R(t)>$)
\cite{Tulapurkar2005, Sankey2006}. This voltage can be expressed as
$V={{1}\over{4}}(R_{AP}-R_{P})I_{\text{rf}}\sin \beta \sin \theta$.
Here $\beta$ is the angle between the free and fixed layers before
applying the rf current and $\theta$ is the precession angle.
$I_\text{rf}$ represents the rf current amplitude. This is a
simplified formula that assumes small angle precession and a
sinusoidal angular dependence of junction resistance between
parallel and antiparallel states. With a perpendicular magnetic
field greater than the free layer's easy-plane anisotropy field, the
free layer magnetization is normal to the surface, while the fixed
layer, which has a larger easy-plane anisotropy field, is still
mainly magnetized in the film plane.  This non-collinear arrangement
of the layer magnetizations ($\beta \lesssim \pi /2 $) enhances the
ST-FMR signal. To further increase the signal (typically in the
sub-$\mu$V range) to noise ratio, we modulate the rf current on and
off at 800 Hz and use a lock-in amplifier to detect the voltage at
this frequency.

\begin{figure}[t]
\includegraphics[width=0.48\textwidth]{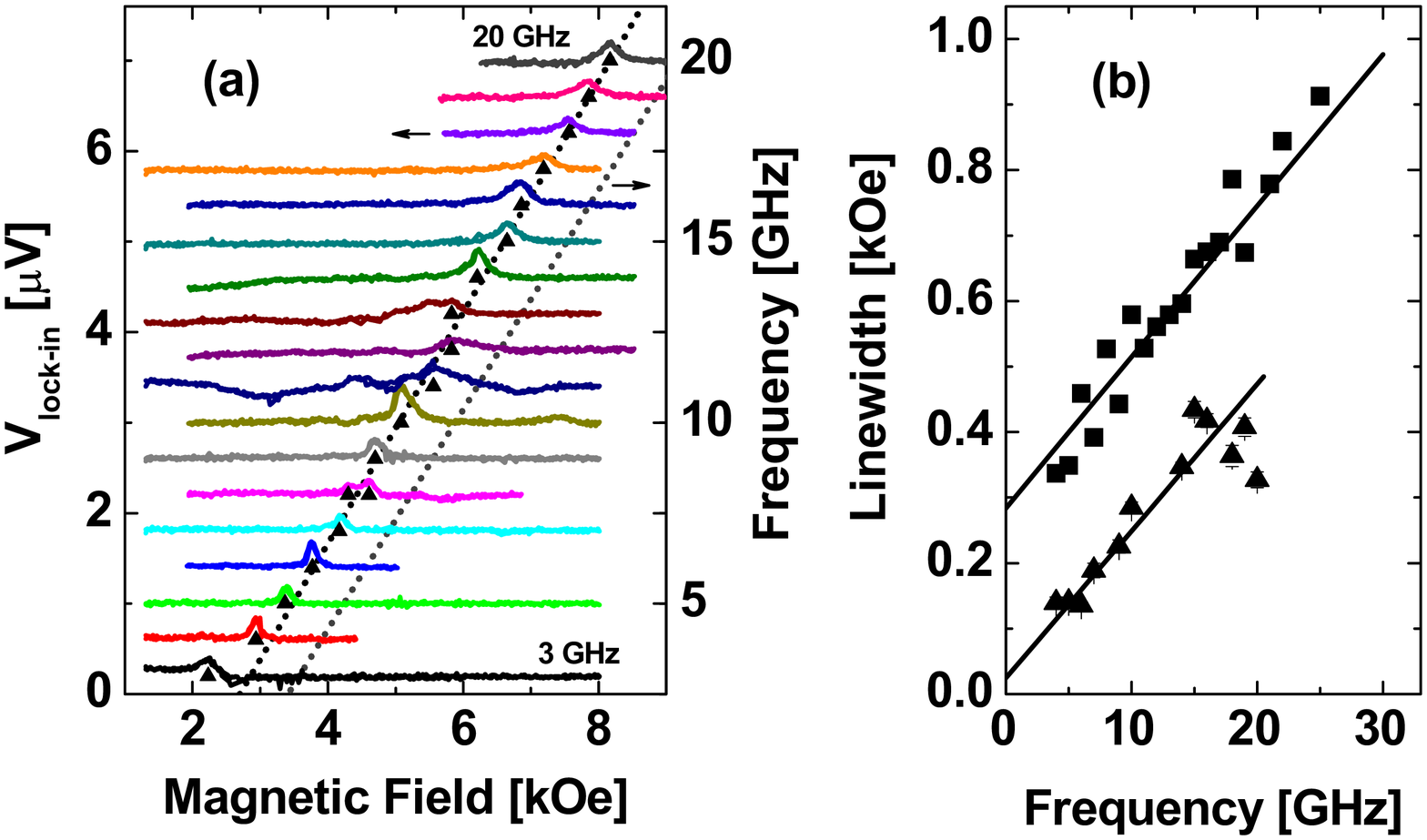}
\caption{(a): Lock-in voltage signal as a function of applied
perpendicular magnetic field at different rf frequencies from 3 up
to 20 GHz in 1 GHz steps. $\blacktriangle$: $H_\text{res}$ of a
50$\times$150 nm$^2$, \textit{t}=0.4 Co/Ni synthetic free layer in a
spin valve; black dashed line: corresponding linear fit; gray dashed
line: a linear fit of $H_\text{res}$ \textit{vs} $f$ of an extended
film with the same Co/Ni layer stack. (b): $\Delta H$ \textit{vs}
$f$ for the spin valve junction ($\blacktriangle$) and the extended
film ($\blacksquare$), together with their corresponding linear
fits.}\label{FMR2}
\end{figure}

ST-FMR measurements were conducted with the circuit shown in Fig.
\ref{DC}(a). Resonance lines under low amplitude rf current at zero
dc current and different rf frequencies $f$ are plotted in Fig.
\ref{FMR2}(a) versus perpendicular magnetic field. Different
frequencies (3$\sim$20 GHz in 1 GHz steps) are plotted with each
adjacent curve offset by 0.4 $\mu$V. The voltage signals are shown
on the left vertical axis. From the peak height $V_\text{peak}$ and
$I_\text{rf}$, we estimate the precession angle to be
$\sim$4$^{\circ}$. We verified that this set of data was taken in a
linear response regime with $V_\text{peak}/I_\text{rf}^2$
independent of $I_\text{rf}$. These data are typical of all
junctions with \textit{t}=0.4. However, much broader resonance peaks
and multiple peaks were found on samples with \textit{t}=0.1, 0.2
and 0.3. This is likely associated with the excitation of higher
order spin wave modes, but is not presently understood. Therefore,
data analysis and discussion mainly focus on samples with
\textit{t}=0.4.

We also measured resonance lines on an extended film with the same
Co/Ni synthetic layer stack sandwiched between 10 nm Cu on each
side. These measurements were conducted with a traditional rf field
driven FMR using a flip-chip method \cite{Beaujour2007}. Broader
resonance peaks were not found in extended films with
\textit{t}=0.1, 0.2, and 0.3.

The resonance field $H_\text{res}$ of the Co/Ni element in the spin
valve increases linearly with $f$ above 4 GHz, as shown in Fig.
\ref{FMR2}(a) ($\blacktriangle$ symbols). At lower frequencies, the
free layer magnetization tilts into the plane, leading to a lower
resonance field. A linear fit of $H_\text{res}$ \textit{vs} $f$ of
the extended film is also plotted with a gray dashed line (to the
right of the $\blacktriangle$ symbols) in Fig. \ref{FMR2}(a). A
linear relationship between $f$ and $H_\text{res}$ in extended
magnetic films is expected when the magnetization is normal to the
film surface ${{h}\over{\mu_{B}}}f=g(H_\text{res}-4\pi
M_\text{eff})$ \cite{Kittel}. Here $g$ is the Land\'{e} $g$ factor
and the easy-plane anisotropy is $4\pi M_{\text{eff}}=4\pi
M_{\text{s}}-H_\text{P}$, where $M_{\text{s}}$ and $H_\text{P}$
represent the saturation magnetization and the perpendicular
anisotropy field. A linear fit of each data set (dashed lines in
Fig. \ref{FMR2}(a)) gives $g$=2.17 and $4\pi M_\text{eff}=2.58$ kOe
for the extended film, and a slightly larger slope (2.28) and a
smaller field-axis intercept (1.92 kOe) for the Co/Ni element
confined in the spin valve. This consistency between data sets
confirms that the main peak of the ST-FMR signal is associated with
the Co/Ni synthetic free layer rather than the other magnetic
layers. The differences are associated with the static dipolar
fields from other magnetic layers and finite size effects on the
spin wave modes, which is discussed in detail in a forthcoming
publication \cite{MMM07}.

\begin{figure}[t]
\includegraphics[width=0.48\textwidth]{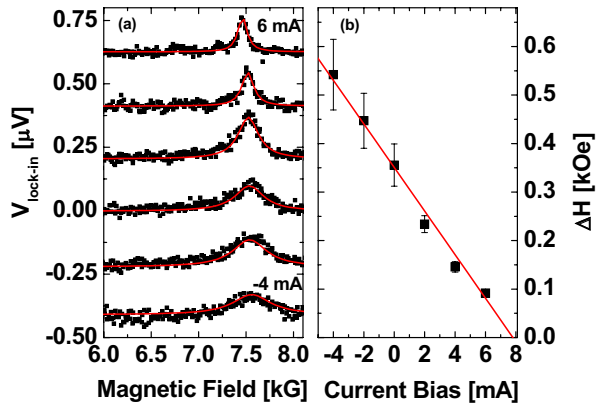}
\caption{(a): ST-FMR signal as a function of applied field at
different dc currents. The rf frequency was set at 18 GHz, and the
rf amplitudes were 595, 595, 470, 470, 315 and 315 $\mu$A
respectively for each dc current from -4 to 6 mA in 2 mA steps. Each
adjacent curve is offset by $0.20~\mu$V. Solid lines are Lorentzian
fits of each data set. (b): $\Delta H$ (full width at half maximum)
\textit{vs} dc current.} \label{FMR1}
\end{figure}

Here we focus on the resonance linewidth $\Delta H$. $\Delta H$
\textit{vs} $f$ at zero dc bias is plotted in Fig. \ref{FMR2}(b).
$\Delta H$ of both the Co/Ni layer in spin valve ($\blacktriangle$)
and the same-stack extended film ($\blacksquare$) increases linearly
with $f$. Linear fits are shown as solid lines in Fig.
\ref{FMR2}(b), and give an intercept and slope:
\begin{eqnarray}
{\Delta H=\Delta H_{0}+{{2\alpha h}\over {g\mu_{B}}}f}\label{eq1}
\end{eqnarray}
where $h$ is the Planck Constant and $\mu_{B}$ is the Bohr Magneton.
The first term $\Delta H_{0}$ describes the inhomogeneous
broadening, and the second term is related to the damping $\alpha$
\cite{Mills}. $\Delta H$ \textit{vs} $f$ of the spin valve and that
of the extended film have a similar slope, implying a similar
damping parameter ($\alpha$=0.036$\pm$0.002 for the extended film
and 0.033$\pm$0.003 for the magnetic layer in the spin valve).
However, the intercepts are quite different: $\Delta
H_{0}$=24$\pm$15 Oe in the spin valve, which is much lower than that
of the extended film, 284$\pm$30 Oe.

When a dc current bias is applied to a spin-value, there is an
additional spin transfer torque that modifies the free layer's
effective damping, $\alpha_\text{eff}=\alpha(1-{{I}\over{I_{c}}})$,
where $I$ is the dc current. Thus $\alpha_\text{eff}$ decreases with
increasing positive current up to a critical current $I_{c}$, that
defines the threshold for magnetic excitation of the free layer. The
critical current in the Slonczewski model \cite{Slonczewski1996} is
given by $I_c= {{2e}\over{\hbar P}}{{\alpha M_\text{s} V}\over{\cos
\beta}}(H-4\pi M_\text{eff})$, where $P$ is the spin polarization
factor and $V$ is the volume of the magnetic element. So with a dc
bias, $\Delta H$-$\Delta H_{0}$ =${{2\alpha
hf}\over{g\mu_{B}}}(1-{{I}\over{I_{c}}})$, and therefore at fixed
frequency $\Delta H$-$\Delta H_{0}$ depends linearly on current and
goes to zero at the critical current. We plot resonance lines of the
spin valve with $f=18$ GHz at different dc currents from -4 to 6 mA
in 2 mA steps in Fig. \ref{FMR1}(a). $\Delta H$ \textit{vs} dc
current bias is shown in Fig. \ref{FMR1}(b). The intercept of
$\Delta H$ \textit{vs} I is 7.8 mA. Inclusion of $\Delta H_{0}$
decreases the intercept by no more than 0.2 mA, because $\Delta
H_{0}$ is small compared to the linewidth at the dc currents
studied. Critical currents determined for $f=$10, 14 and 18 GHz are
plotted in Fig. \ref{DC}(d), and agree well with those obtained from
the I-V measurements. Further, $I_{c}$ is quantitatively consistent
with the Slonczewski model taking a spin polarization factor P
$\sim$0.3.

The frequency independent term $\Delta H_{0}$ originates from film
inhomogeneities: roughness, polycrystalline structure, as well as
defects. The scale of the inhomogeneities is likely the film grain
size, 5$\sim$10 nm. In a simple model, fluctuations in
$H_\text{res}$ from grain to grain result in an inhomogeneously
broadened resonance line \cite{Hurdequint2002}. However, it is
likely that the exchange coupling between grains is important to a
detailed understanding of the linewidth \cite{McMichael2003}.

The free layer in the nanostructured device contains at most a few
hundred grains, therefore one expects less inhomogeneity than that
in extended film. More importantly, the lateral magnetic confinement
results in strongly varying internal field in the plane of the
nanostructure that lifts the degeneracy between different spin wave
modes. Numerical and analytical calculations of normal modes in the
Co/Ni rectangular element are presented and compared with our ST-FMR
data in Ref. \cite{MMM07}. The separation between them is more than
the inhomogeneous broadening $\Delta H_{0}$ in the extended film,
therefore we expect that the linewidth measured on an individual
mode is close to its intrinsic value (the term proportional to $f$
in Eq. \ref{eq1}) \cite{deLoubens2007}. The remaining inhomogeneous
broadening in the nanostructure may be attributed to the
quasi-degeneracy subsisting between some very closely spaced modes
resulting from film inhomogeneities.

In summary, ST-FMR has been used to study the magnetic properties of
Co/Ni synthetic layers with perpendicular anisotropy in spin valves.
The ST-FMR resonance lines were compared with those of traditional
FMR on same-stack extended film. The damping of the ST-device free
layer is essentially the same as that of an unpatterned film and the
critical currents determined from the ST-FMR homogeneous linewidth
are in agreement with those of quasistatic I-V measurements.

This research is supported by NSF-DMR-0706322 and an NYU-Research
Challenge Fund award.

\end{document}